\documentclass[11pt]{article}
\usepackage{jheppub}
\usepackage{pstool}
\pdfoutput=1
\usepackage{color}
\usepackage{float}
\usepackage{array}
\usepackage{amssymb}
\usepackage{amsmath}
\usepackage{amsthm}
\usepackage{graphicx}
\usepackage{caption}
\usepackage[labelsep=quad]{subcaption}
\usepackage{epstopdf}
\usepackage{placeins}
	
\usepackage{epsfig}

\def\pb[#1,#2]{\{#1, #2\}}
\def\deb[#1,#2]{[#1,#2]_{\text{D.B.}}}

\def\ep{\epsilon}

\def\Or[#1]{{\text{O}}\left({#1}\right)}
\def\dotl[#1,#2]{\left\langle #1,\, #2 \right\rangle}
\def\dotlb[#1,#2]{\left\langle #1,\, #2 \right\rangle}
\def\dotlm[#1,#2]{\left[ #1,\, #2 \right]}
\def\dotp[#1,#2]{(\vect{#1} \cdot\vect{#2})}
\def\aff[#1,#2]{\hat{#1}(#2)}

\def\n4sym{{\cal N}=4 SYM}
\def\>{\rangle}
\def\<{\langle}
\def\weight[#1,#2,#3]{\{(#1),#2,#3\}}
\def\ads[#1]{$\text{AdS}_{#1}$}

\newcommand{\be}{\begin{equation}}
\newcommand{\ee}{\end{equation}}
\newcommand{\ba}{\begin{align}}
\newcommand{\ea}{\end{align}}
\newcommand{\bs}{\begin{split}}
	\def\sess\end{split}
\newcommand{\vect}[1]{{\boldsymbol{#1}}}

\def \bea {\begin{eqnarray}}
\def \eea {\end{eqnarray}}
\def \bea* {\begin{eqnarray*}}
	\def \eea* {\end{eqnarray*}}

\def \be {\begin{equation}}
\def \ee {\end{equation}}
\def \bes {\begin{equation*}}
\def \ees {\end{equation*}}

\newtheorem{theorem}{Theorem}

\def\ep{\epsilon}

\title{Conformal Blocks and Bilocal Vertex Operator Transition Amplitudes}
\author[a]{Gideon Vos,}
\emailAdd{vos@fzu.cz}
\affiliation[a]{Central European Institute for Cosmology, \\ FZU, Na Slovance 1999/2, 182 21 Prague 8, Czech Republic\\}

\date{}

\abstract{We revisit the construction of the 2d conformal blocks of primary operator four-point functions as bilocal vertex operator correlators. We find an additional interpretation as a path integral over the reparametrizations of an intermediate cylinder. As a consequence we bridge the gap between the K\"ahler quantization of virasoro coadjoint orbits, $SL(2,\mathbb{R})$ Chern-Simons theory and the reparametrization formalism of 2d CFT that has made an appearance in recent literature.}
\begin{document}
	\maketitle

\section{Introduction}	
Very quickly after the construction of two-dimensional conformal field theory it became known that these models are closely intertwined with the analytic geometry of Riemann surfaces. In particular, it was realized that the conformal blocks of the 2d CFT can be thought of as sections of a Hermitian projective line bundle of the underlying Riemann surfaces \cite{Friedan:1986ua}. In this short letter we will consider a relatively simple special case. The identity block contribution to the four-point function of two pairs of identical primary operators $\langle O_V O_V O_W O_W\rangle$ on a Riemann sphere background. We will write the identity block contribution as a path integral over deformations of an intermediate cylinder that connects to open disks on which the primary operators are made to live. Requiring that all three disconnected components subsequently weld back into a single surface is a subtle non-trivial procedure that automatically ends up promoting the tangent vector fields of Teichm\"uller spaces of the individual components into a single global complex-valued field $\ep(z,\bar{z})$. This field will turn out to be identical to the reparamatrization expansion of \cite{Cotler:2018zff,Nguyen:2020jqp}, the soft mode of \cite{Haehl:2018izb} and the shadow field of \cite{Haehl:2019eae}. Hence, this note will provide a analytic geometrical derivation of the reparametrization mode formalism that has been studied recently \cite{Cotler:2018zff,Haehl:2019eae,Haehl:2018izb,Anous:2020vtw,Nguyen:2020jqp}

We will view the identity conformal block of the four-point function as a quantum transition amplitude of wavefunctions that are sections of a holomorphic line bundle on the Virasoro coadjoint orbit $\overline{Diff}/SL(2,\mathbb{R})$. This will be reviewed by constructing the holomorphic line bundle on a closely related space, the Teichm\"uller space of the open disk $\mathcal{T}(\Delta)$ \cite{Verlinde:2021kgt}, some basics of which are briefly reviewed in the appendix\footnote{The space $\overline{Diff}/SL(2,\mathbb{R})$ is in reality just a single leaf of a foliation of $\mathcal{T}(\Delta)$ \cite{NagVerjovsky}, but since the symplectic form on $\overline{Diff}/SL(2,\mathbb{R})$ induces the Weil-Peterson K\"ahler form on $\mathcal{T}(\Delta)$ for the purposes of this article these spaces can be thought of as essentially the same.}.

\begin{figure}
	\centering
		\includegraphics[scale=0.3]{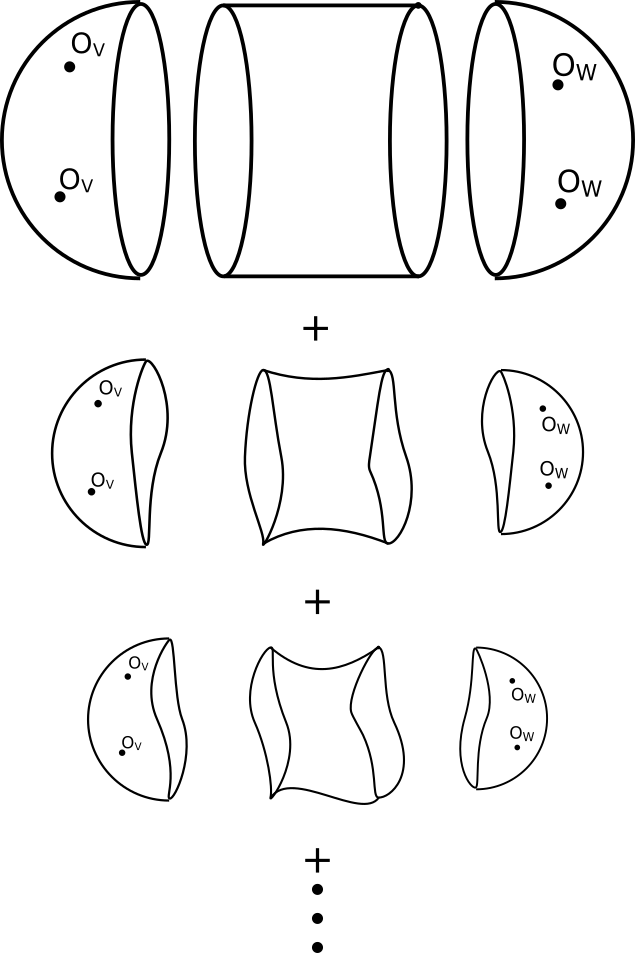}
	\caption{The sum over deformed intermediate cylinders that reproduces the identity block}
	\label{CylinderSum}
\end{figure}

The result is a Hermitean combination that is an integral over Teichm\"uller space modulated by an integration density that is naturally given by the exponent of the K\"ahler potential in order to ensure invariance under K\"ahler transformations. The point of this paper is to show that this can be naturally recast into a visually appealing picture of that of an integral over the reparametrizations of an intermediate cylinder geometry fig.\ref{CylinderSum}. Requiring that the resulting three disconnected surfaces conformally weld back together into a single Riemann surface naturally reproduces the integrand of the mentioned transition amplitude. It is oft-stated in the literature \cite{Verlinde:1989hv} without detail that one cut open a Riemann surface and place CFT states on the boundaries on both sides of the cut. In this paper we will explicitly describe, through means of the conformal welding theorem and the equivalence of A-model and B-model Teichm\"uller space (see appendix A for an explanation of these terms), a geometrical protocol that ensures that both sides off the cut contain the same CFT information.

There are two particularly satisfying upshots to this approach, for one it makes the rather abstract K\"ahler quantization very visible, secondly of the presented analysis none of it inherently requires taking a large central charge limit\footnote{It should be immediately mentioned that this statement comes with a disclaimer, the analysis as presented does not require a central charge limit. If one would actually want to go and compute the identity block in practical terms one would likely have to introduce a large central charge limit.}. As a smaller upshot, we manage to avoid common but somewhat imprecise terminology (`attaching a CFT state to a Riemann surface boundary') through means of an explicit geometrical protocol.

Throughout this note the fundamental underpinning of the above connection between the four-point conformal blocks and the Wilson line transition amplitudes of Chern-Simons theory is the equivalence between the Teichm\"uller space of the open disk and the space of diffeomorphisms of the boundary circle of that disk. It is tempting to think that in the specific case of AdS$_3$/CFT$_2$ the mathematical underpinning of the gauge/gravity duality is the possibility to reduce the information of Teichm\"uller space to boundary data.

\section{K\"ahler quantization}
The K\"ahler quantization approach to CFT is not the conventional approach in contemporary CFT literature, therefore in this short section some basic results that are critical for remainder of this paper will be reviewed. 

In schematic form, the vacuum conformal block $V_I$ contribution to the four-point function
\begin{equation}
\langle O_V O_V O_W O_W \rangle,
\end{equation}
is given by
\begin{equation}
\mathcal{F}_I = \sum_{\{ p \}} \; \langle O_V O_V L_{-n} ... L_{-m}|0\rangle \, \langle 0| L_m ...L_n O_W O_W \rangle.
\end{equation}
Here ${p}$ indicates the set of ordered partitions of the positive integers. This expression is schematic as we are ignoring the orthonormalization required to make this set of states into an orthonormal basis. For pragmatic purposes assume the set of orthonormal basis vectors on the Virasoro vacuum representation $\{ \tilde{L}_{n,k}|0\rangle\}$, for a representation build out of $|0\rangle$ this can be obtained by removing all combination containing $L_{-1}$ and subsequently normalizing. The following expression is now exact though relatively academic in nature
\begin{equation}
\mathcal{F} = \sum_{n,k} \langle O_V O_V \tilde{L}_{n,k}|0\rangle \langle 0|\tilde{L}^{\dagger}_{n,k} O_W O_W \rangle.
\label{block}
\end{equation}
The two individual factors in the sum can be interpreted as the wavefunction $\Psi_W(\{p\})$ and Hermitian conjugate wavefunction $\bar{\Psi}_V(\{p\})$ of the respective states $O_V O_V|0\rangle$ and $O_W O_W|0\rangle$ on the Virasoro vacuum representation. The wavefunctions are explicitly given by
\begin{equation}
\Psi_W (n,k) = \langle 0| \tilde{L}^{\dagger}_{n,k}  O_W O_W|0\rangle, \;\;\; \bar{\Psi}_V (n,k) = \langle O_V O_V \tilde{L}_{n,k}|0\rangle.
\end{equation}
These are the wavefunctions of the state on a Hilbert space which is a unitary representation of a Lie group. In the case of a compact the Lie group the wavefunctions can be represented as sections of a holomorphic line bundle over $G/H$ where $H$ is the Cartan subgroup of $H$. In the case of the non-compact Virasoro group the situation is much more subtle but this situation has been worked out in the past, the wavefunctions are now sections of a holomorphic line bundle over the Virasoro coadjoint orbit. In particular, since we are interested in the highest weight representation with highest weight state $|0\rangle$ the relevant coadjoint orbit is $\overline{Diff}(S^1)/SL(2,\mathbb{R})$, which will hence be referred to as the vacuum orbit\footnote{the overline implies a central extension, i.e. elements of the orbit are pairings of a quadratic differential and a real number. We introduce the notation in advance for consistency later on.}. \cite{segal, lazutkin, Witten:1987ty}  

The above sum over wavefunctions now takes the form of an integral over the Virasoro coadjoint orbit
\begin{equation}
\mathcal{F}_I(V,W) = \sum_{n,k} \langle \Psi_V(n,k) | \Psi_W(n,k)\rangle = \int_{\overline{Diff}(S^1)/SL(2,\mathbb{R})} dF \; e^{-K(F,\bar{F})} \bar{\Psi}_V(\bar{F}) \Psi_W(F).
\end{equation}
The functions $F$ and $\bar{F}$ are group coordinates, they are trajectories on the space of diffeomorphisms of the circle $S^1$, $F(\tau, \phi)$ where $\bar{F}(\tau, \phi)$ with periodicity conditions $F(\tau, \phi +2\pi) = F(\tau, \phi) +2\pi$ and $\bar{F}(\tau, \phi +2\pi) = \bar{F}(\tau, \phi) +2\pi$. In principle, the functions $F(\tau, \phi)$ and $\bar{F}(\tau,\phi)$ when Fourier decomposed in their angular coordinate only respectively depend on the positive frequencies and the negative frequencies. 

Here the function $K(F,\bar{F})$ is the K\"ahler potential of the Kirilov-Kostant form of the $\overline{Diff}(S^1)/SL(2,\mathbb{R})$ orbit
\begin{equation}
K= \frac{c}{24\pi} \int d\phi d\tau \, \frac{\dot{F}}{F'}\left(\frac{F'''}{F'}-2\frac{F''^2}{F'^2}\right).
\end{equation}
The prefactor $c$ is the CFT central charge, $\dot{F}$ indicates a derivative with respect to $\tau$, $F'$ indicates a derivative with respect to $\phi$.

The sections transform as $\Psi_W(n,k) \rightarrow e^{\alpha(F)}\Psi_W(n,k)$ and $\bar{\Psi}_V(n,k)\rightarrow e^{\bar{\alpha}\bar{F}}\bar{\Psi}_V(n,k)$ respectively under K\"ahler transformations $K\rightarrow K +\alpha(F)+\bar{\alpha}(\bar{F})$. It can be shown that the combination above is the unique Hermitian combination that is invariant under these gauge transformations \cite{Verlinde:1989hv}.


\subsection{Constructing the holomorphic line bundle}
In order to costruct the holomorphic line bundle over the appropriate Virasoro coadjoint orbit it will be easier instead to consider an equivalent K\"ahler manifold, the Teichm\"uller space of the open unit disk $\mathcal{T}(\Delta)$. Some details of Teichm\"uller space are discussed in appendix A, but the basic idea is that $\mathcal{T}(\Delta)$ is the universal cover of the moduli space of inequivalent complex structures that the open disk $\Delta$ can be endowed with. The catch is to determine how two complex structures are inequivalent. These complex structures can be obtained from a reference coordinate frame by a quasiconformal mapping of the disk to itself $\Delta \rightarrow \Delta$, a mapping is non-trivial if it induces a non-trivial boundary structure on the circle that makes up the boundary of the disk. Two complex structures are inequivalent if their two associated induced mappings at the boundary are inequivalent. Hence the Teichm\"uller space is given by the space of diffeomorphisms of the circle, in particular for the open disk $\mathcal{T}(\Delta)$ this space is modded by the isometry group of the disk $SU(1,1) \cong SL(2,\mathbb{R})$, hence $\mathcal{T}(\Delta) \cong Diff(S^1)/SL(2,\mathbb{R})$.\\

\noindent For every inequivalent complex structure a local metric can be obtained of the form $ds^2 = e^{\phi}|dz + \mu(z,\bar{z})d\bar{z}|^2$ with Beltrami differential field $\mu(z,\bar{z})$. Locally, on any coordinate patch one can always construct a coordinate transformation that brings the metric to the form $ds^2 = e^{\phi}dz d\bar{z}$ such that the metric has constant curvature. Hence the space of inequivalent complex structures corresponds to the space of inequivalent constant curvature metrics.

Following \cite{Verlinde:1989ua}, we can parametrize the space of constant curavature metrics by a set of two zweibein 1-forms $e^{+}, e^{-}$ and a spin connection field $\omega$. These fields are subject to the following constraints:
\begin{eqnarray}
d\omega + e^{+} \wedge e^{-} = 0,\\
d e^{+} - \omega \wedge e^{+} = 0,\\
d e^{-} + \omega \wedge e^{-} = 0.
\end{eqnarray}
The first constraint states that the metric $ds^2 = e^{+} \otimes e^{-}$ has constant curvature everywhere and the last two constraints ensure that the metric has zero torsion. The space of constant curvature metrics is now a constrained phase space manifold coordinatized by the two-component one-forms $e^{+}, e^{-}, \omega$ with a symplectic form given by
\begin{equation}
\Omega = \frac{k}{2\pi} \int_{\Delta} \delta \omega(e^{+},e^{-}) \wedge \delta \omega(e^+,e^-) - \Lambda \delta e^{+} \wedge \delta e^{-},
\end{equation}
where $\Lambda$ is the constant curvature and $k$ is an arbitrary constant.

We can proceed to quantize this phase space by constructing the holomorphic line bundle whose base space consist of half the coordinates of the phase space. One can use a particularly convenient choice of polarization on the phase space manifold
\begin{equation}
\omega_{\bar{z}} = i\frac{4\pi}{k} \frac{\delta}{\delta \omega_z(z)}, \;\; e^{-}_z = \frac{4\pi}{k}\frac{\delta}{\delta e^{+}_{\bar{z}}(z)}, \;\; e^{-}_{\bar{z}} = -i\frac{4\pi}{k} \frac{\delta}{\delta e^{+}_z(z)}.
\end{equation}
The wave functionals $\Psi[e^{+}_z,e^{+}_{\bar{z}},\omega_z]$ will be the functionals of the fields $e^{+}_z$, $e^{+}_{\bar{z}}$, $\omega_{z}$. The resulting space of wave functionals is to large as we have not accounted for the constraints above. 
reparametrizing the zweibeins through
\begin{equation}
e^{+} = e^{\phi}\left(dz + \mu d\bar{z}\right), \;\; e^{-}=e^{\bar{\phi}}\left( d\bar{z} + \bar{\mu}dz\right),
\end{equation}
and using two out of three constraints to eliminate the $\phi$ and $\omega$ dependence (see \cite{Verlinde:1989ua} for details) leaves one last remaining constraint operator constraint for the wavefunctional $\Psi[\mu]$ as a functional of only $\mu$. This operator constraint is given by 
\begin{equation}
\left(\bar{\partial} - \mu \partial - 2 \partial \mu \right) \frac{\delta}{\delta \mu} \Psi[\mu] + \frac{ic}{24\pi} \partial^3 \mu \Psi[\mu] = 0.
\end{equation}
The disk can be dressed with $n$ additional punctures at the points $z_i$ with associated scaling weights $h_i$. In this case the above expression is modified to
\begin{equation}
\left(\bar{\partial} - \mu \partial - 2 \partial \mu \right) \frac{\delta}{\delta \mu} \Psi[\mu, z_i] =  \left(-\frac{c}{24\pi} \partial^3 \mu + \sum_i^n (h_i \partial \delta(z-z_i) + \delta(z-z_i)\partial_{z_i}\right) \Psi[\mu, z_i]
\label{virasoroward}
\end{equation}
This can be recognized as the Virasoro Ward identity. What is not particularly trivial is that this expression for infinitesimal reparametrizations can be solved for global reparametrizations of the surface. This solution is given by
\begin{equation}
\Psi[\mu, z_i] = e^{-\frac{c}{24\pi} \int d^2z\, \frac{\bar{\partial} F}{\partial F} \left(\frac{\partial^3 F}{\partial F} - 2\left(\frac{\partial^2 F}{\partial F}\right)^2\right) - \frac{c}{24\pi} \int d^2 z\, \mu S[F,z]} \prod_i^n (\partial f(z_i))^{h_i} \langle O_1 (f(z_1)) ... O_n(f(z_n))\rangle,
\label{globalward}
\end{equation}
here $F(z,\bar{z})$ is the formal inverse $F(f(z,\bar{z}),\bar{z})=z$ of the solution to the Beltrami equation $f(z,\bar{z})$
\begin{equation}
\mu(z,\bar{z}) = \frac{\bar{\partial}f}{\partial f}.
\end{equation}
In addition $O_j(z_j)$ is a conformal primary operator with scaling weight $h_j$ and $S[F,z]$ indicates the Schwarzian derivative with respect to $z$, i.e.
\begin{equation}
S[F,z] = \frac{\partial^3 F}{\partial F} - \frac{3}{2} \left(\frac{\partial^2 F}{\partial F}\right)^2.
\end{equation}
Hence it is the reparametrized primary correlation functions form the wave-functionals of the Teichm\"uller space and by extension the Virasoro coadjoint orbit. Due to the coupling of the CFT stress tensor to the metric, the left-hand side of the expression \eqref{globalward} can alternatively be written as
\begin{equation}
\Psi[\mu, z_i] = \langle e^{-\int d^2z\, \mu T} O(z_1)...O(z_n)\rangle.
\end{equation}
It is equation \eqref{globalward} that will play a fundamental role in the subsequent analysis. This expression has the interpretation of a transformation rule for conformal correlators under the chiral transformation $(z,\bar{z}) \rightarrow (f(z,\bar{z}),\bar{z})$. Note that if $f(z,\bar{z})$ were holomorphic that the Alekseev-Shatashvili action would vanish and the expression would take the familiar form of the transformation rule of conformal correlators under general conformal transformations.

\subsection{Wave functionals of the four-point function}
To be entirely explicit, in the situation under consideration in this paper we are dealing with wave functionals on the open disk with two primary operator insertions. As our base manifold is the open disk we can construct a set of global conformal initial coordinates $ds^2 = dz d\bar{z}$ that cover the entire disk, as a result we can set the initial Beltrami differential to zero. The resulting wavefunctional is
\begin{equation}
\Psi_W[\mu] = e^{-\frac{c}{24\pi} \int_{\Delta} d^2z\, \frac{\bar{\partial} F}{\partial F} \left(\frac{\partial^3 F}{\partial F} - 2\left(\frac{\partial^2 F}{\partial F}\right)^2\right)} \left(\frac{\partial f(z_3) \partial f(z_4)}{(f(z_3)-f(z_4))^2}\right)^{h_W}.
\end{equation}
In terms of $\mathcal{T}(\Delta)$ the vacuum block contribution to the four-point function takes the form
\begin{align}
&\mathcal{F}_I(V,W) = \int_{\mathcal{T}(\Delta)} Df\, e^{-\frac{c}{24\pi} \int_{cyl} d\phi d\tau \, \frac{\dot{F}}{F'}\left(\frac{F'''}{F'}-2\frac{F''^2}{F'^2}\right)} \nonumber \\
& e^{-\frac{c}{24\pi} \int_{\Delta} d^2z\, \frac{\bar{\partial} F_V}{\partial F_V} \left(\frac{\partial^3 F_V}{\partial F_V} - 2\left(\frac{\partial^2 F_V}{\partial F_v}\right)^2\right)} \left(\frac{\partial f_V(z_1) \partial f_V(z_2)}{(f_V(z_1)-f_V(z_2))^2}\right)^{h_V} \nonumber \\
& \times e^{-\frac{c}{24\pi} \int_{\Delta} d^2z\, \frac{\bar{\partial} F_W}{\partial F_W} \left(\frac{\partial^3 F_W}{\partial F_W} - 2\left(\frac{\partial^2 F_W}{\partial F_W}\right)^2\right)} \left(\frac{\partial f_W(z_3) \partial f_W(z_4)}{(f_W(z_3)-f_W(z_4))^2}\right)^{h_W}
\label{blockintegral}
\end{align}
We are now integrating over the Teichm\"uller space of the open disk parametrized by the inequivalent quasiconformal maps of the disk. Here the integral over the cylinder takes into account the contribution of the K\"ahler potential. The weakness of this expression is that it is not a priori obvious \textit{which} quasiconformal mapping $f_V, f_W$ of the disk corresponds to which trajectory over the vacuum orbit that the K\"ahler potential takes its value over. It is precisely this ambiguity that we will resolve in the subsequent text through means of conformal welding and the equivalence of A-model and B-model Teichm\"uller space.

\section{A geometric construction of the bilocal vertex operator correlator}
We will now attempt to write the previous expression for the vacuum conformal block in such a way that we obtain an appealing visual picture that connects with the formulation of the identity block as a two-point function of bilocal vertex operators. In the process we will end up naturally replicating the form of \eqref{blockintegral}, in addition, by demanding a fairly obvious constraint (that our disconnected surfaces weld back together again into a single manifold) we resolve quasiconformal map ambiguity of \eqref{blockintegral} discussed above.

\begin{figure}
	\centering
		\includegraphics[scale=0.4]{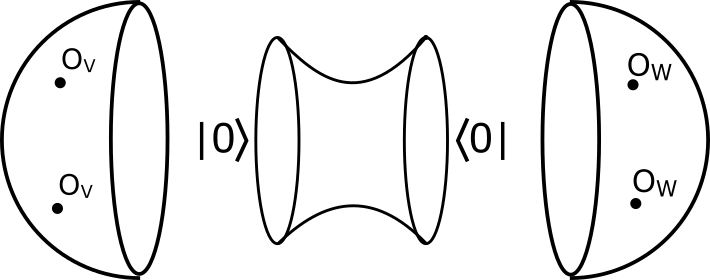}
	\caption{Split up the sphere on which the four primary operators reside. Fixing the CFT states living on the boundaries of the intermediate geometry to the vacuum state has the effect of inserting a $|0\rangle \langle 0|$ projector into the four-point correlator.}
	\label{NodeInsertion}
\end{figure}

In order to accomplish this, draw a non-contractable cycle on the Riemann sphere with the four primary operator insertions in such a way that the cycle divides up the sphere in a part with just $O_V$ operators and $O_W$ operators. Now contract this cycle down such that the sphere gets pinched to an intermediate node geometry. Demanding that the CFT state on the boundaries of the resulting node are the CFT vacuum states is not an invariant operation. It is essentially a visual way of inserting the vacuum projector $|0\rangle \langle 0|$, see fig. \ref{NodeInsertion}, which has the effect of factorizing the correlator.
\begin{equation}
\langle O_V O_V O_W O_W \rangle \rightarrow \langle O_V O_V |0\rangle \langle 0|0\rangle \langle 0| O_W O_W\rangle.
\label{split}
\end{equation}
The vacuum-to-vacuum transition amplitude has been added explicitly to emphasize the point that this is product of three individual factors.

We will focus on the geometry of the intermediate node for a moment. The near node geometry has the topology of an annulus and the moduli space of this topology can be parametrized by a single complex parameter $q$, this number parametrizes the separation distance and relative twist of the two boundaries of the annulus. The way this parametrization comes through is by taking two coordinate disks $\left\{z_{1,2}\;\, | \;\; |z_{1,2}|<1\right\}$. Take the complex parameter $|q|<1$, remove the subdisks with radius $\sqrt{|q|}$ from the two unit disks and identify the two resulting annuli at the inner boundaries. As stated in \cite{Friedan:1986ua} this resulting geometry can be coordinatized by a single coordinate $z$ given in terms of the original two coordinates $z_{1,2}$ by
\begin{equation}
z = 
\begin{cases}
\frac{\sqrt{q}}{z_2} \;\;\;\; \text{if} \, \sqrt{|q|}<|z| \leq 1, \\
\frac{z_1}{\sqrt{q}} \;\;\;\; \text{if} \, 1 \leq |z| < \frac{1}{\sqrt{|q|}}.
\end{cases}
\end{equation}
In the limit $q\rightarrow 0$ the coordinate $z$ starts to cover the entire Riemann sphere with the exception of punctures at the origin and infinity. It is this limit where the annulus geometry pinches off to a node. The coordinate transformation $w = \frac{1}{2\pi i} \log z$ realizes the node as a cylinder. This will be useful later.

\subsection{Reparametrizing the intermediate node}
Once we have applied the logarithmic map to turn the node geometry into a cylinder, the result is a cylinder with CFT vacuum states attached to the boundaries. By the state-operator map we can contract these boundary circles down to points and insert identity operators at these points, resulting in the simple sphere partition function $\langle 0|0\rangle$ which can be assumed to be normalized to unity. We can coordinatize the Riemann sphere with the usual isothermal coordinates $z, \bar{z}$ such that the line element takes the form $ds^2 = dz d\bar{z}$. The situation get more interesting though when we perform a chiral reparametrization of the sphere of the form $(z,\bar{z}) \rightarrow (f(z,\bar{z}), \bar{z})$. In which case the sphere partition function becomes the Polyakov action \cite{Polyakov} of 2d induced gravity
\begin{equation}
W[f] = \frac{c}{24\pi} \int d^2z \, \frac{\bar{\partial}f}{\partial f} \partial^2 \ln (\partial f).
\end{equation}  
By writing the complex coordinates in terms of cylindrical coordinates $z=\tau + i \phi$, $\bar{z} = \tau - i\phi$. We can perform a change of fields by defining $F(\tau, \phi)$ as the formal inverse of the above transformation, i.e. $F(f(\tau,\phi),\phi)=\tau$ in which case the action above becomes the Alekseev-Shatashvili action \cite{Alekseev:1988ce,AS90}
\begin{equation}
W[F] = \frac{c}{24\pi} \int  d\tau d\phi \, \frac{\dot{F}}{F'}\left(\frac{F'''}{F'}-\frac{2}{3}\left(\frac{F''}{F'}\right)^2\right),
\end{equation}
which can be recognized as the K\"ahler potential of the Virasoro coadjoint orbit $\overline{Diff}(S^1)/SL(2,\mathbb{R})$. At least, in the case where the function $F(\tau,\phi)$ satisfies the periodicity condition $F(\tau, \phi+2\pi) = F(\tau,\phi) +2\pi$ and monotonicity condition $F'(\tau,\phi)>0$.

This mapping has the effect of deforming the cylinder $S^1 \times [0,\beta]$ by attaching a continuous family of diffeomorphism of the circle along the axis of the cylinder. The resulting partition function of the surface is weighted by the Alekseev-Shatashvili action above. We will assume the for now rather arbitrary periodic boundary condition that the diffeomorphisms match on both sides of the cylinder, i.e. $F(0,\phi)=F(\beta,\phi)$. This will be justified later by putting the action on a torus though heuristically it is because we want the in- and out-state of the projector \eqref{block} to match when we compute the conformal block.

\subsection{Conformal welding}
\begin{figure}
	\centering
		\includegraphics[scale=0.2]{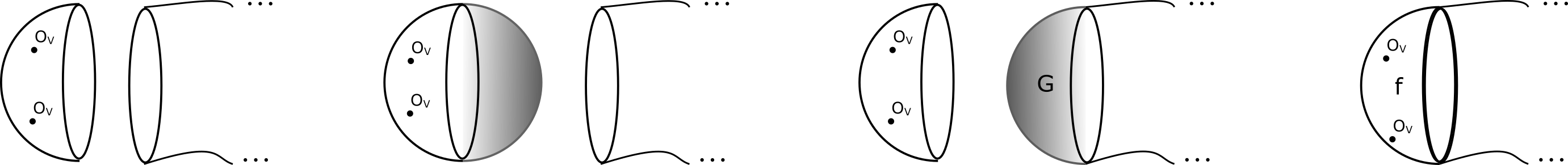}
	\caption{A graphical representation of the conformal welding procedure. We extend the surface by sewing a cap on the $S^1$ boundary, we use the conformal welding theorem to weld this cap disk to the cylinder by deforming it with a biholomorphic map $g$. We interpret the biholomorphic map on this disk as the holomorphic map on the complement region that parametrizes an element of the B-model Teichm\"uller space of the disk. We extend this B-model holomorphic function into the two-punctured disk as a quasi-conformal map parametrized by the map $f$.}
	\label{conformalwelding}
\end{figure}

After deforming the intermediate surface it is no longer obvious that the three individual subsurfaces `click' back together again in order to make a single full surface. In order to complete the argument we will require both models of Teichm\"uller space reviewed in appendix A, what we will do is take one of the disk spaces with operator insertions and complete it to a full Riemann sphere by sewing a cap to the unit disk. This cap we will in turn weld to one end of the deformed cylinder. As we will see, the conformal welding theorem states that this can be accomplished by a biholomorphic map of the disk. We will then interpret the cap as the complement region of the unit disk in the B-model and interpret the holomorphic welding map as an element of B-model Teichm\"uller space of the two-punctured disk. Finally we will interpret the resulting quasisymmetric map that forms the conformal weld as a mode parametrizing the A-model Teichm\"uller space. This results in a glued surface (see figure \ref{conformalwelding}). We will subsequently interpret the wavefunctionals of the previous section as the correlator of two primary operators dressed by a quasi-conformal map. This quasiconformal map acts as point on a classical phase space whose quantization yields the vacuum represenation of the Virasoro, hence this protocol will yield the usual CFT intuition of attaching a CFT state to the boundary of a surface.

The essential tool to glue all three surfaces together is a modified version of the conformal welding theorem of two disks
\begin{theorem}
If $h:S^1 \rightarrow S^1$ is quasisymmetric then there exist biholomorphic maps $F$ and $G$ from $\Delta$ and $\hat{C}\setminus\Delta$ into complementary Jordan domains $\Omega$ and $\Omega^*$ of $\hat{C}$ such that $F^{-1}\circ G|_{S^1} = h$. Moreover, the Jordan curve separating $\Omega$ and $\Omega^{*}$ is a quasicircle.
\end{theorem}
The exact phrasing was taken from \cite{radnell}, that paper also discusses the procedure that generalizes the statement above to more general topologies than disks. For clarification, quasicircle means the image of a circle under a quasisymmetric map.

This essentially states that as long as the seam is given by a quasisymmetric map of the circle that the there exists a biholomorphic map on both our punctured disk and our cylinder such that the total surface uplifts the complex structure of the subsurfaces. The theorem states that on the full glued surface the seam where they are joined together will be a quasicircle. The logic will be the following, since the deformation of the middle cylinder surface acts as a quasisymmetric map on the boundary circles of the cylinder we can suspect that the resulting quasicircle acts as the conformal weld. 

To formalize this intuition, since the theorem above holds for any quasisymmetric map we can identify it with the quasisymmetric map that takes the boundary of the undeformed cylinder to the boundary of the deformed cylinder. As a result, since the cylinder is already `in place' the biholomorphism $F$ that maps the cylinder to the Jordan domain coincides with the identity map at the boundary, i.e. $F|_{S^1}=I$. Inserting this information in the expression $F^{-1}\circ G|_{S^1} = h$, forces the restriction $G|_{S^1}$ to coincide with the quasisymmetric map $h$. 

We now interpret the biholomorphic map of the disk as the solution to the Beltrami equation outside the two-punctured disk in the B-model. Hence, it is the extension of a quasi-conformal map on the two-punctured disk. Constructing the Beltrami differential associated to this quasi-conformal map and extending it to the sphere as in the A-model gives us a quasisymmetric map at the edge of the operator insertion disk that coincides with the conformal weld, the whole surface connects at the seam.

\subsection{The conformal block and the analytic geometry of 2d CFT}
Each of the three individual factors of \eqref{split} take a different form on the welded surface. Just to reiterate, we deformed the intermediate cylinder and subsequently deformed the two end-caps with operator insertions such that the total surface could be glued back together again. We will indicate by $f_i(z,\bar{z})$, $i=1,2$ the two quasi-conformal maps that deform the disks where the operators live and the reparametrization of the intermediate cylinder by $f(z,\bar{z})$. In this case, by applying the global Ward identity \eqref{globalward} the three factors of \eqref{split} take the form
\begin{align}
& e^{-\frac{c}{24\pi} \int_{cyl} d\phi d\tau \, \frac{\dot{F}}{F'}\left(\frac{F'''}{F'}-2\frac{F''^2}{F'^2}\right)} \nonumber \\
& e^{-\frac{c}{24\pi} \int_{\Delta} d^2z\, \frac{\bar{\partial} F_1}{\partial F_1} \left(\frac{\partial^3 F_1}{\partial F_1} - 2\left(\frac{\partial^2 F_1}{\partial F_1}\right)^2\right)} \left(\frac{\partial f_1(z_1) \partial f_1(z_2)}{(f_1(z_1)-f_1(z_2))^2}\right)^{h_V} \nonumber \\
& \times e^{-\frac{c}{24\pi} \int_{\Delta} d^2z\, \frac{\bar{\partial} F_2}{\partial F_2} \left(\frac{\partial^3 F_2}{\partial F_2} - 2\left(\frac{\partial^2 F_2}{\partial F_2}\right)^2\right)} \left(\frac{\partial f_2(z_3) \partial f_2(z_4)}{(f_2(z_3)-f_2(z_4))^2}\right)^{h_W}.
\label{integrand}
\end{align}
Here the functions $F_i(z,\bar{z})$ are once again defined as the inverses of $f_i(z,\bar{z})$, i.e. $F_i(f(z,\bar{z}),\bar{z}) =z$. This expression can be immediately recognized as the integrand of \eqref{blockintegral}, but now with the exact properties of the quasi-conformal maps specified. The result is now clear, the expression of the conformal block \eqref{blockintegral} can be interpreted as a path integral over all reparametrizations of an intermediate cylinder under the conditions that disks where the operators live are simultaneously deformed such that the three surfaces can be glued back together.

\subsubsection{Conformal block in terms of the node modulus series expansion}
This short section functions as a side note that might that casts light from another angle on the same general construction. The spirit of the above construction can be traced back into the past, a similar approach to conformal correlators was investigated in \cite{Friedan:1986ua}. In which the authors discuss a certain factorization property of a projective line bundle on the compactification divisor of punctured Riemann surfaces. In particular they follow the same initial procedure
\begin{itemize}
\item We draw a non-contractible cycle on the Riemann surface.
\item We pinch together the cycle to create a new Riemann surface with a node.
\end{itemize}
It was argued that the conformal four-point punction of primary operators\footnote{Their results apply to more general conformal correlators and even more general background topologies, but we will focus on the part that coincides with correlator under consideration.} has a power series expansion in terms of the modular parameter $q$ of the intermediate node geometry
\begin{equation}
Z(m, \bar{m}) = \sum_{\phi} q^{h_{\phi}}\bar{q}^{\bar{h}_{\phi}} Z_{\phi}(m_D, \bar{m}_D).
\label{partitionexpansion}
\end{equation}
This expression requires some explanation. The parameters $m, \bar{m}$ represent the moduli of the surface before introducing the node. The sum over $\phi$ represents the sum over all states of definite scaling weight of the CFT (both primary and descendent). The sum coefficients $Z_\phi$ are functions of the moduli $m_D, \bar{m}_D$, the moduli of the compactification divisor, i.e. the moduli space of surfaces \textit{with} nodes. In our simple case the moduli $m,\bar{m}$ of the initial surface are just the locations and scaling weights of the primary operators. This is again a  formal way of inserting a complete set of states, we insert a node geometry and sum over all possible CFT states that can live on the boundaries of the node.

In the limit $q\rightarrow 0$ the above sum is dominated by a single term, the vacuum state contribution. In this case the coefficient $Z_{0}(m_D, \bar{m}_D)$ was found to simply the product of correlators of the disconnected surfaces, i.e. $Z_{0}(m_D, \bar{m}_D) = \langle O_V O_V \rangle \langle O_W O_W \rangle$. Hence this coefficient is known, we will argue in this short section that the result of the previous section has an interpretation in terms of the continuum limit of the sum \eqref{partitionexpansion}. We will argue that the coeffients of the sum restricted to descendants of the vacuum in the continuum limit is given by the reparametrized two-point functions that allow for conformal welding.  

In order to do this take the partition function of the reparametrized intermediate cylinder
\begin{equation}
W[F] = -\frac{c}{24\pi} \int  d\tau d\phi \, \frac{\dot{F}}{F'}\left(\frac{F'''}{F'}-\frac{2}{3}\left(\frac{F''}{F'}\right)^2\right),
\end{equation}
We will now path integrate this expression against the symplectic measure $DF = df/f'$ \cite{Alekseev:1988ce,AS90,Cotler:2018zff,Mertens:2017mtv}. We will once again assume the periodic boundary condition $F(0,\phi)=F(\beta,\phi)$. This puts the path integral on a torus with period $\beta$ hence the expectation value is given
\begin{equation}
\int_{\text{periodic}} [Df] e^{-\frac{c}{24 \pi} \frac{\dot{F}}{F'}\left(\frac{F'''}{F'}-\frac{2}{3}\left(\frac{F''}{F'}\right)^2\right)} = \underset{V_I}{\text{Tr}} \; e^{-\beta H}.
\end{equation}
The trace is taken over the Virasoro vacuum module, as this is the quantum hilbert space of the action above. The stress tensor components of the action above are known \cite{Alekseev:1988ce}, in particular the $T^{00}$ component is given by
\begin{equation}
T^{00} = -\frac{c}{24 \pi} \; \{f, \theta \},
\end{equation}
resulting in the Hamiltonian
\begin{equation}
H = -\frac{c}{24 \pi} \int_0^{2\pi} d\theta \; \{f, \theta \}
\end{equation}
Which is part of a conventional basis of Virasoro charges of the action
\begin{equation}
L_n = -\frac{c}{24 \pi} \int_0^{2\pi} d\theta \; \{f,\theta\} e^{in\theta}.
\end{equation} 
By reintroducing the modular parameter $q$ through $q= e^{\beta}$ the above path integral takes the form
\begin{equation}
\int_{\text{periodic}} [Df] e^{-\frac{c}{24 \pi} \frac{\dot{F}}{F'}\left(\frac{F'''}{F'}-\frac{2}{3}\left(\frac{F''}{F'}\right)^2\right)} = \underset{V_I}{\text{Tr}} \; q^{-L_0}.
\end{equation}
Which is indeed the sum of basis functions of the series expansion \eqref{partitionexpansion} restricted to the vacuum representation. We therefore propose that these expressions are one and the same and additionally similar in the physical intuition that underpins them. It should be emphasized that this similarity is just heuristic and meant to provide some additional intuition to the rather abstract K\"ahler quantization construction.

\section{The bilocal vertex transition function}
The above expression \eqref{integrand} still contains an ambiguity, there are many quasi-conformal maps that induce the correct data on the boundary of the disks. In appendix A we discussed another parametrization of points of Teichm\"uller space, we can construct a point by exponentiating a tangent vector attached to the origin of $\mathcal{T}(\Delta)$. Another upshot is that while the above expression for constructing the integrand by finding the appropriate conformal weldings corresponding to $f$ is well-defined it is not very practical, as we will see the field redefinition below will include the appropriate welding in a very straightforward manner.

Constructing a tangent vector corresponds to solving an infinitesimal version of the Beltrami equation with infinitessimal Beltrami differential $\delta \mu$ 
\begin{equation}
\frac{\bar{\partial}f}{\partial f} = \delta \mu
\end{equation}
This expression van be solved perturbatively through von Neumann series \cite{Donaldson} and the result is given by
\begin{equation}
f=z + \ep + \ep \partial \ep - \bar{\partial}^{-1}(\ep \bar{\partial}\partial \ep) +...
\end{equation}
where $\ep$ is related to $\delta \mu$ through $\delta \mu = \bar{\partial} \ep$ and the inverse derivatives are defined through an integral relation
\begin{equation}
\partial^{-1}g(z,\bar{z}) = \int d^2 z'\, \frac{g(z',\bar{z}')}{\bar{z}-\bar{z}'}.
\end{equation}
The idea is to perform a change of field variables away from the quasiconformal maps $f, f_i$ in favor of the respective fields $\ep_{\text{cyl}}$, $\ep_i$ instead. As an aside, the von Neumann series expansion of the infinitesimal Beltrami equation seems to exactly reproduce the reparamatrization expansion of \cite{Cotler:2018zff,Nguyen:2020jqp}, the soft mode of \cite{Haehl:2018izb} and the shadow field of \cite{Haehl:2019eae}, suggesting that this is the geometrical underpinning at the foundation of those constructions. By inverting $F$ and inserting the series expansion above we find the following perturbative expansion for the Alekseev-Shatashvili action
\begin{equation}
\frac{c}{24\pi} \int  d^2 z \, \frac{\bar{\partial}F}{\partial F}\left(\frac{\partial^3 F}{\partial F}-\frac{2}{3}\left(\frac{\partial^2 F}{\partial F}\right)^2\right) \rightarrow \frac{c}{24\pi} \int  d^2 z \, \bar{\partial} \ep \partial^3 \ep+... 
\end{equation}
With this expression of the geometrical action in terms of the $\ep$-fields the above expression for the identity block takes the form
\begin{align}
& \Psi_I(z_i) =\int [D\epsilon_{\text{cyl}}] e^{-\frac{c}{24 \pi} \int_{cyl} \bar{\partial} \ep_{\text{cyl}} \partial^3 \ep_{\text{cyl}}+... } e^{-\frac{c}{24 \pi} \int_{\Delta^2_2} \bar{\partial} \ep_1 \partial^3 \ep_1 +...} e^{-\frac{c}{24 \pi} \int_{\Delta^2_2} \bar{\partial} \ep_2 \partial^3 \ep_2 +...} \\ 
& \times \left(\frac{(1+\partial \ep_1(z_1)+...)(1+\partial \ep_1(z_2)+...)}{\left(z_1 - z_2 +\ep_1(z_1) - \ep_1(z_2)\right)^2}\right)^{h_V}  \left(\frac{(1+\partial \ep_2(z_3)+...)(1+\partial \ep_2(z_4)+...)}{\left(z_3 - z_4 +\ep_2(z_3) - \ep_2(z_4)\right)^2}\right)^{h_W} \\
& =\int [D\epsilon_{cyl}] e^{-\frac{c}{24 \pi} \int_{\Sigma} \bar{\partial} \ep_{\Sigma} \partial^3 \ep_{\Sigma} +...}\left(\frac{(1+\partial \ep_{\Sigma}(z_1)+...)(1+\partial \ep_{\Sigma}(z_2)+...)}{\left(z_1 - z_2 +\ep_{\Sigma}(z_1) - \ep_{\Sigma}(z_2)\right)^2}\right)^{h_V} \\
& \times \left(\frac{(1+\partial \ep_{\Sigma}(z_3)+...)(1+\partial \ep_{\Sigma}(z_4)+...)}{\left(z_3 - z_4 +\ep_{\Sigma}(z_3) - \ep_{\Sigma}(z_4)\right)^2}\right)^{h_W}
\end{align}
Note that demanding that the surfaces weld together properly means that the original diffeomorphisms $f_i$ and $f$ have to coincide at their common boundaries. This is a heavy constraint for the fields $\ep_{cyl}$, and the two $\ep_i$ fields, the series expansion for $f$ above reveals that not only do the values of $\ep$-fields have to coincide on the conformal welds all the derivatives of the field $\ep_i, \ep_{\text{cyl}}$ have to coincide as well. This means that the fields $\ep_{i}$ uniquely extend to $\Delta_i$. Because the domains of the fields combine smoothly we can combine the fields $\ep_{cyl}$ and $\ep_{i}$ together into a single field $\ep_{\Sigma}$ that lives on the entire welded surface. In addition since the extension into the disks is unique  we can extend the path integral without overcounting
\begin{align}
& \Psi_I(z_i) =\int [D\epsilon_{\Sigma}] e^{-\frac{c}{24 \pi} \int_{\Sigma} \bar{\partial} \ep_{\Sigma} \partial^3 \ep_{\Sigma} +...}\left(\frac{(1+\partial \ep_{\Sigma}(z_1)+...)(1+\partial \ep_{\Sigma}(z_2)+...)}{\left(z_1 - z_2 +\ep_{\Sigma}(z_1) - \ep_{\Sigma}(z_2)\right)^2}\right)^{h_V} \\
& \times \left(\frac{(1+\partial \ep_{\Sigma}(z_3)+...)(1+\partial \ep_{\Sigma}(z_4)+...)}{\left(z_3 - z_4 +\ep_{\Sigma}(z_3) - \ep_{\Sigma}(z_4)\right)^2}\right)^{h_W}
\end{align}
This reveals the role of the $\epsilon_{\Sigma}$ field, it can be thought of as a ``gauge choice" on the disks that resolves the quasi-conformal map ambiguity mentioned above.

The resulting expression for the identity block is a perturbative quantum field theory of the mode $\epsilon_{\Sigma}$ with a propagator that is fixed by the quadratic term of the expansion of the Alekseev-Shatashvili action. Hence the resulting identity block can be computed through standard perturbative quantum field theory techniques in the regime where the central charge $c$ is large and standard Feynman diagram techniques can be applied. In fact the global field $\ep(z,\bar{z})$ can now be recognized as the reparamatrization expansion of \cite{Cotler:2018zff,Nguyen:2020jqp}, the soft mode of \cite{Haehl:2018izb} and the shadow field of \cite{Haehl:2019eae}. The result is similar to identical to expressions obtained in the past, notably \cite{Cotler:2018zff,Nguyen:2020jqp,Haehl:2018izb,Haehl:2019eae,Anous:2020vtw} suggesting that these results can all be obtained as consequences of the analytic geometry underlying two-dimensional CFT.

\FloatBarrier

\section{Discussion}
In this short letter we discussed a procedure where we pinch off an intermediate node geometry from the four-punctured sphere with two sets of identical punctures. Subsequently we path integrate over all reparametrization of the intermediate node and deform the two disconnected end caps by conformally welding them back together into a single surface at every point of the integration space. We find that this procedure exactly replicates the in-to-out state transition amplitude in the K\"ahler quantization approach to 2d CFT. In the process, by `gauge fixing' a remaining ambiguity in the space of quasiconformal mappings we constructed a quantum field theory over a global reparametrization field $\ep(z,\bar{z})$ that exactly replicates the reparamatrization expansion of \cite{Cotler:2018zff,Nguyen:2020jqp}, the soft mode of \cite{Haehl:2018izb} and the shadow field of \cite{Haehl:2019eae}, hence providing an independent analytical geometry interpretation to those varying approaches.

\begin{figure}
	\centering
		\includegraphics[scale=0.4]{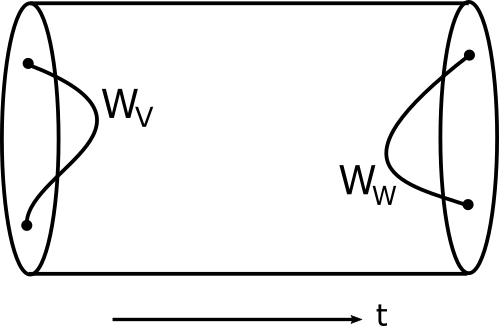}
	\caption{The above identity block can be recast as the transition amplitude of two Wilson lines in a single $SL(2,\mathbb{R})$ Chern-Simons theory}
	\label{WilsonLineTransitionAmplitude}
\end{figure}

It should be noted that the construction of the wavefunctionals \cite{Verlinde:1989ua} that was reviewed in section 2 was originally used to demonstrate that the states of $SL(2,\mathbb{R})$ Chern-Simons theory are given by the conformal blocks of 2d CFT. As a result the four-point vacuum block has a dual interpretation as a full quantum transition function of two Wilson lines fig. \ref{WilsonLineTransitionAmplitude}. Though not new, it is satisfying to see a `path integral over geometry' contribute to a 2d correlation function without the strict requirement of a semi-classical large central charge limit.

\begin{figure}[h]
	\centering
		\includegraphics[scale=0.4]{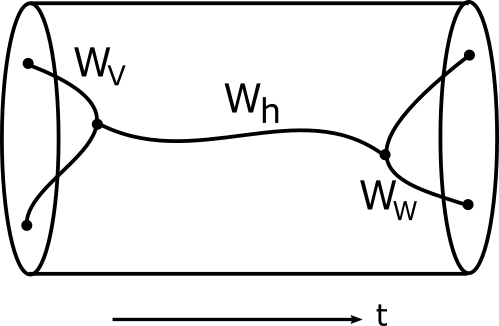}
	\caption{By modifying the assumed highest weight state at the boundaries of the cylinder and by extension the Virasoro coadjoint orbit under consideration one finds that the Wilson lines fuse to an intermediate line with a holonomy fixed by the constant representative of the orbit.}
	\label{WilsonLineFusion}
\end{figure}

A brief note on higher conformal blocks where the intermediating representation is a highest weight representation with a heighest weight primary state that is not the vacuum. Note that in order to construct the vacuum block, as we did above, our starting point was an intermediate cylinder with vacuum states living on the boundaries, i.e. a sphere by the state-operator map. We could have inserted two identical states at the boundaries that are primary but not the vacuum, i.e. $|h\rangle$. In this case the additional term in the exponent of the global Ward identity \eqref{globalward} would not be trivial and the result would be that the geometrical action would be one of the Alekseev-Shatashvili actions of the orbits $\overline{Diff}/S^1$. As these orbits are related to the Teichm\"uller space of the punctured disk \cite{Verlinde:2021kgt}, we would have to weld a cap to the cylinder that is the complement region of a punctured disk. The result would have a Chern-Simons theory interpretation where the two Wilson lines are fused together to an intermediate Wilson line of fixed holonomy fig. \ref{WilsonLineFusion}.

\section*{Acknowledgements}
The author of this note has benefitted greatly from various discussions with various people. Firstly I would like to thank Kevin Nguyen for collaborating on an earlier version of this project. In addition, I would like to thank Souvik Banerjee, Felix Haehl, Ondrej Hulik, Tomas Prochazka and Joris Raeymaekers for discussion on this work. This work was supported by the Grant Agency of the Czech Republic
under the grant EXPRO 20-25775X

\appendix

\section{Basics of Teichm\"uller theory}
In this section we will cover some Teichm\"uller theory preliminaries which will be required to put all three disconnected surfaces back together again. First we will review some basic general properties of Teichm\"uller theory. The Teichm\"uller space of the open disk quotiented by a Fuchsian group $\Delta/ \Gamma$ is given by the space of inequivalent complex structures. Here inequivalent means that one cannot map from one complex structure to another with a diffeomorphism that does not leave the boundary of the disk untouched \footnote{Note that in this aspect there is a small conflict between the string theory literature and the CFT or Chern-Simons theory literature adopted here. In string theory there is no such constraint at the boundary. Our definition of Teichm\"uller space maintains a distinction between `large' diffeomorphisms and `small' diffeomorphisms as in the 3d gravity literature. We consider two points in Teichm\"uller space equivalent only if they can be transformed into each other without introducing gravitational hair at the boundary. W mention this specifically because some string theorists might be confused by the claim the Teichm\"uller space of the open disk is non-trivial. See also the distinction between Teichm\"uller space and reduced Teichm\"uller space in \cite{radnell}}. Equivalently this is the space of representations of the group $\Gamma$ up to an equivalence relation, in this case two representations are considered equivalent if one can be obtained from the other by conjugating all group generators by the same matrix.  

The Teichm\"uller space is parametrized by the Beltrami differentials $\mu(z,\bar{z})$, under elements of the discrete group $\gamma \in \Gamma$ these are constrained by the transformation rule \\
$\mu (\gamma z) \overline{\gamma'(z)}/\gamma'(z) = \mu(z)$. In addition we assume that $|\mu|<1$ everywhere on the disk. Here, the group elements $\gamma$ are realized as M\"obius transformations. The Beltrami differentials as is form an excessive set of moduli as two distinct $\mu_i$ can be equivalent. We now require to know how to discern if two Beltrami differentials $\mu_1, \mu_2$ are equivalent. For this we require the Beltrami equation
\begin{equation}
\mu = \frac{\bar{\partial} f}{\partial f}.
\end{equation}   
The condition that $|\mu|<1$ everywhere ensures that $f(z,\bar{z})$ is an orientation-preserving map of the unit disk to itself and falls in the class of quasiconformal mappings. There are essentially two ways of extending this equation from the open disk to the complex plane, the first one is the so-called A-model where we extend $\mu$ by Schwarz reflection around the unit circle
\begin{equation}
\tilde{\mu} =
\begin{cases}
\overline{\mu(1/\bar{z})}\frac{z^2}{\bar{z}^2}, \;\; z\in \hat{\mathbb{C}}\backslash \Delta, \\
\mu(z) \;\;\;\;\;\;\;\;\;\;\; z\in \Delta.
\end{cases}.
\end{equation}
For the resulting extended Beltrami equation $\mu_1$ and $\mu_2$ are equivalent if and only if their associated quasiconformal maps coincide on the boundary
\begin{equation}
\mu_1\sim \mu_2 \; \Leftrightarrow \; f_1|_{S^1} = f_2|_{S^1}.
\end{equation}
The point is now the following. We can invert the logic above and parametrize the Teichm\"uller space by the space of mappings $f:S^1\rightarrow S^1$ that are the boundary restrictions of the quasiconformal maps, these are generally dubbed quasisymmetric maps. It should be noted that the there is still some residual redundancy, in order to truly parametrize Teichm\"uller space in terms of quasisymmetric maps one has to quotient out a particular set of equivalent transformation. In the cases relevant to us this will generally be either $SU(1,1) \cong SL(2,\mathbb{R})$ or the group of rigid rotations $U(1)$. 

Now take as an element of Teichm\"uller space a point on a trajectory parametrized by a flow line from a tangent vector $\delta \mu$ attached to some reference point. The Beltrami equation can be solved by von Neumann series for such an infinitesimal vector \cite{Donaldson}. The result is 
\begin{equation}
f= z + \bar{\partial}^{-1} \delta\mu + \bar{\partial}^{-1} \delta \mu \partial \bar{\partial}^{-1}\mu + ...
\end{equation}
Locally we can construct a local vector field $\bar{\partial} \ep =\delta \mu$, in which case $f$ locally turns into
\begin{equation}
f=z + \ep + \ep \partial \ep - \bar{\partial}^{-1}(\ep \bar{\partial}\partial \ep) +...
\end{equation} 
By taking into account all higher-order terms we effectively `exponentiate' the tangent vector. Thus, $\ep$ parametrizes the same information as the Beltrami differential. It is tempting to think that all knowledge of the equivalence class of $\ep$ is contained in its restriction to the unit circle, this would be inaccurate though. $f|_{S^1}$ depends on all derivatives of $\ep$ constrained to the unit circle and therefore all information of $\ep$ on at least an open set is required. Note that $\ep$ is only locally a vector field and cannot necessarily be globally patched together, if it can be globally patched together in such a way that it transforms as a vector field under the holonomy transformation $\Gamma$ then $f$ is automatically trivial on the unit circle \cite{Ahlfors}.

The alternative approach is the B-model, in this case the Beltrami differential is simply set to zero outside of the unit disk
\begin{equation}
\tilde{\mu} =
\begin{cases}
0, \;\;\;\;\;\;\;\; z\in \hat{\mathbb{C}}\backslash \Delta, \\
\mu(z), \;\;\; z\in \Delta.
\end{cases}.
\end{equation}
The solution to the Beltrami equation on the whole Riemann sphere outside the disk is a holomorphic function. Unlike the A-model, in this model two Beltrami differentials are equivalent if the holomorphic extensions outside the disk coincide, i.e.
\begin{equation}
\mu_1\sim \mu_2 \; \Leftrightarrow \; f_1(z)|_{\hat{\mathbb{C}}\backslash \Delta} = f_2(z)|_{\hat{\mathbb{C}}\backslash \Delta}.
\end{equation}
Hence in this case elements of Teichm\"uller space are alternatively parametrized by holomorphic functions on the complement of the unit disk rather than by quasisymmetric maps. Both these models are equivalent through conformal welding \cite{Nag}.

\end{document}